\documentclass[prd, aps, superscriptaddress, preprintnumbers, twocolumn, floatfix, nofootinbib]{revtex4}
\pdfoutput=1

\usepackage{amsfonts}
\usepackage{amsmath}
\usepackage{amssymb}
\usepackage{bm}
\usepackage{dcolumn}
\usepackage{graphicx}
\usepackage[latin1]{inputenc}
\usepackage{latexsym}
\usepackage{rotating}
\usepackage{hyperref}
\usepackage{graphicx}
\usepackage{color}

\newcommand\be{\begin{equation}}
\newcommand\bea{\begin{eqnarray}}
\newcommand\ee{\end{equation}}
\newcommand\eea{\end{eqnarray}}

\renewcommand{\d}{{\mathrm{d}}}

\renewcommand{\(}{\left(}
\renewcommand{\)}{\right)}
\newcommand{\nn}{\nonumber}
\newcommand{\Mpl}{M_{\textrm{Pl}}}

\def\doi{http://doi.org}

\def\d{\mathrm{d}}

\begin{document}

\title{Entanglement entropy of cosmological perturbations for S-brane Ekpyrosis}

\author{Suddhasattwa Brahma}
\email{suddhasattwa.brahma@gmail.com}
\affiliation{Department of Physics, McGill University, Montr\'{e}al, QC, H3A 2T8, Canada}

\author{Robert Brandenberger}
\email{rhb@physics.mcgill.ca}
\affiliation{Department of Physics, McGill University, Montr\'{e}al, QC, H3A 2T8, Canada}

\author{Ziwei Wang}
\email{ziwei.wang@mail.mcgill.ca}
\affiliation{Department of Physics, McGill University, Montr\'{e}al, QC, H3A 2T8, Canada}

\date{\today}

\begin{abstract}
 
 We calculate the entanglement entropy of scalar perturbations due to gravitational non-linearities present in any model of canonically-coupled, single-field ekpyrosis. Specifically, we focus on a recent model of improved ekpyrosis which is able to generate a scale-invariant power spectrum of curvature perturbations and gravitational waves as well as have a non-singular bounce due to an S-brane at the end of ekpyrotic contraction. By requiring that the entanglement entropy remians subdominant to the thermal entropy produced during reheating, we get an upper bound on the energy scale of the bounce.
 
\end{abstract}

\maketitle

\section{Introduction}

In a recent paper \cite{us} we studied the momentum space entanglement entropy between sub- and super-Hubble modes generated by the intrinsic gravitational nonlinearities. We applied the formalism to inflationary cosmology and found that the entanglement entropy is a growing function of time since the phase space of super-Hubble modes is increasing. By demanding that the entanglement entropy remain smaller than the thermal entropy after reheating, we found an upper bound on the duration of inflation which is consistent with the constraint \cite{TCC2} coming from the {\it trans-Planckian censorship conjecture} \cite{TCC1}. In this paper we will apply the formalism developed in \cite{us} to the new version of the Ekpyrotic scenario recently proposed by two of us \cite{NewEkp1}. Once again, the phase space of super-Hubble modes is an increasing function of time, and hence the entanglement entropy will be increasing. By demanding that the entanglement entropy at the end of the phase of Ekpyrotic contraction be smaller than the thermal entropy after the bounce, we find an upper bound on the energy scale of the bounce.

The Ekpyrotic scenario \cite{Ekp} is a promising alternative to cosmological inflation. The scenario is based on the assumption that the cosmological scale factor $a(t)$ is decreasing very slowly
\be
a(t) \, \sim \, t^p \,\,\,\, {\rm{with}} \,\,\, 0 < p \ll 1 \,
\ee
as a function of physical time $t$, where $t < 0$ in the contracting period. In the context of Einstein gravity, this time dependence of the scale factor can be obtained if matter is dominated by a scalar field with a negative exponential potential. Such potentials are ubiquitous in string theory compactifications (see e.g. \cite{Baumann}). A contracting universe with $w > 1$ leads to a dilution of spatial curvature and of anisotropies \cite{Wesley}. It is also an attractor in initial space \cite{Ijjas}. These are significant advantages compared to other cosmological models with a contracting phase (see e.g. \cite{Peter} for a review of bounce scenarios).

Recently, two of us \cite{NewEkp1} proposed a new version of the Ekpyrotic scenario in which an S-brane (see e.g. \cite{Sbrane}) arising at string densities mediates the non-singular transition between contraction and expansion. If the S-brane has zero shear, then roughly scale-invariant spectra of both cosmological perturbations and gravitational waves are generated \cite{NewEkp2}, with two consistency relations between the four basic cosmological observables (the amplitudes and tilts of the scalar and tensor spectra). The decay of the S-brane after the bounce naturally generates \cite{NewEkp3} the radiation of the post-bounce universe. In this paper we will consider entropy constraints on this new scenario.

In the following section we briefly review the equations which describe the phase of Ekpyrotic contraction. In Section III we summarize the framework for computing momentum space entanglement between sub- and super-Hubble modes developed in \cite{us}.  We then turn to the computation of the entanglement entropy at the end of the phase of Ekpyrotic contraction, as a function of the Hubble parameter at that time, and derive the resulting upper bound on the energy scale of the bounce by demanding consistency with the second law of thermodynamics for the full time evolution of the model.

We close this section with some general remarks. We already introduced the cosmological scale factor $a(t)$ which yields the Hubble expansion rate
\be
H(t) \, = \, \frac{{\dot{a}}}{a} \, ,
\ee
where an overdot represents a derivative with respect to time. The inverse of $H$ is the Hubble radius. The Hubble radius plays a crucial role concerning cosmological perturbations (see e.g. \cite{MFB, RHBrev} for reviews): on length scales smaller than the Hubble radius fluctuations oscillate. On super-Hubble scales the oscillations freeze out, the fluctuations can be squeezed and classicalize \cite{Kiefer}. Both during an expanding inflationary cosmology and during Ekpyrotic scenario the Hubble radius decreases in comoving coordinates, and in both scenarios a natural assumption is that fluctuations start as quantum vacuum perturbations on sub-Hubble scales. The modes we have access to are the ones which have been able to classicalize. Hence, it is natural to divide the total Hilbert space of fluctuation modes into sub- and super-Hubble modes, and to study the entanglement entropy between these two sets of modes. Since the phase space of super-Hubble modes is increasing as a function of time, the resulting entanglement entropy will also be increasing.
In order to be consistent with the entropy in the expanding radiation phase, the entanglement entropy at the end of the contracting phase cannot be too large. It is the resulting constraint on the model which we will here explore.

We will be using natural units in which the speed of light, Planck's constant and Boltzmann's constant are set to $1$. We denote the energy density by $\rho$ and the pressure by $P$, keeping the symbol $p$ for the exponent appearing in the expression for the time evolution of the scale factor (to be consistent with the usual notation in papers on the Ekpyrotic scenario).

\section{Description of S-brane ekpyrosis}

In the usual realizations of the Ekpyrotic scenario, the  slow contraction phase is driven by a scalar field with negative exponential potential \cite{Ekp}
\begin{equation}
 S \, = \, \int dt d^3x\big( - \frac{1}{2} \partial^\mu\varphi\partial_\mu\varphi + V_0e^{-\sqrt{\frac{2}{p}}\frac{\varphi}{\Mpl}}\big)\, ,
\end{equation}
where $\Mpl$ is the Planck mass.
Since the potential is steep, the field is not slowly rolling and we can find an approximate solution by ignoring the Hubble parameter $H$. This approximation is further justified since during Ekpyrosis the equation of state parameter $w \equiv \frac{P}{\rho} \gg 1$ and hence the contraction of the universe is very slow. Using this approximation, we find that the scalar field evolves according to
\begin{equation}
 \varphi(t) \, = \, \sqrt{2p}\;\Mpl\; \log\left(-\sqrt{\frac{V_0}{p}}\frac{t}{\Mpl}\right)\,.
\end{equation}
Further, using the relation $\dot\varphi^2 = -2\dot{H} \Mpl^2$, we find that
\begin{equation}
 \dot{H} = -\frac{p}{t^2}, ~~~~ H = -\frac{p}{-t}, ~~~~ a = a_0(-t)^{p}\,.
\end{equation}

One can check the self-consistency of our approximation by calculating the equation of state
\begin{eqnarray}
 w \, &=& \, \frac{P}{\rho} = \frac{-\rho+ \dot\varphi^2}{\rho} \\
 &=& \, \frac{-3H^2\text{m}^2_\text{pl} - 2\dot H \text{m}^2_\text{pl}}{3H^2\text{m}^2_\text{pl}}
 = -1 + \frac{2}{3p} \gg 1\,. \nonumber
\end{eqnarray}
From the above, the ``slow-roll'' parameters can be calculated in a straightforward manner. Note that only the first order slow-roll parameter is non-zero and is given by
\begin{equation}
 \epsilon = -\frac{\dot H}{H^2} = \frac{1}{p}, ~~~~\eta \equiv \frac{\dot\epsilon}{H\epsilon} = 0\,.
\end{equation}
It will be beneficial for us later on to rewrite the quantities above in terms of conformal time
\begin{equation}
 \tau(t) \, = \, \int \frac{dt}{a(t)} = -\frac{1}{1-p}(-t)^{1-p}\,.
\end{equation}
The background quantities are given by
\bea
a(\tau) \, &=& \, \tilde a_0(-\tau)^{\frac{p}{1-p}}, \nonumber \\
H(\tau) \, &=& \,  p(1-p)^{-\frac{1}{1-p}}(-\tau)^{-\frac{1}{1-p}} \\
\mathcal H(\tau) \, &\equiv& \,
\frac{da}{ad\tau} \, = \, -\left(\frac{p}{1-p}\right)\frac{1}{\tau}\,. \nonumber
\eea

In \cite{NewEkp1} it was suggested that an S-brane (a space-like brane) will be generated once the background energy density reaches the string scale. The S-brane is an object with vanishing energy density and negative pressure (since it has positive tension). Hence, it is an object which violates the NEC (null energy condition) and allows for the nonsingular transition between the Ekpyrotic contracting phase and a radiation dominated phase of expansion.

\section{States for super-Hubble modes}

In order to study the entanglement entropy due to mode-coupling between cosmological perturbations, we study linear fluctuations about a spatially flat FLRW metric. In longitudinal gauge, the metric can be written as \cite{MFB, RHBrev}
\begin{eqnarray}
 \d s^2 \, = \, - a^2(\tau) \bigl[\d\tau^2(1 + 2 \phi) - (1 + 2 \psi) \d{\bf{x}}^2 \bigr] \,,
\end{eqnarray}
where $\phi$ and $\psi$ are the metric perturbation variables. A particularly useful variable is the curvature fluctuation $\zeta$ in uniform density gauge which is given by
\be
\zeta \, = \, - \psi + \frac{H}{\dot{\rho}} \delta \rho \, ,
\ee
where $\rho$ is the background energy density and $\delta \rho$ is the density perturbation.
If matter is a single field minimally coupled to Einstein gravity (as is the case in the Ekpyrotic scenario), the variable in terms of which the action for fluctuations has canonical form is
\be
v(\bm x, \tau) \, \equiv \, z(\tau)\zeta(\bm x,\tau) \, ,
\ee
where 
\be
z(\tau) \, = \,  \sqrt{2\epsilon}\; a\; \Mpl c_s^{-1} \, ,
\ee
where $c_s^2$ is the speed of sound squared of the matter source. We can expand this field in Fourier modes and canonically quantize it, introducing mode creation and annihilation operators $a_k^{\dagger}$ and $a_k$, respectively.

The Hamiltonian derived from the quadratic action in momentum space is organized as follows (see e.g. \cite{Hamilt})
\bea
H_2 \, &=& \,  \frac{1}{2}\int \frac{d^3k}{(2\pi)^3}\left[c_s k (a_{\bm k} a_{\bm k}^{\dagger} + a_{-\bm k}a_{-\bm k}^{\dagger} )\right]  \nonumber \\
&&- \frac{1}{2}\int
\frac{d^3k}{(2\pi)^3}\left[i\left(\frac{z'}{z}\right)(a_{\bm k}a_{-\bm k}-a_{\bm k}^{\dagger}a_{-\bm k}^{\dagger})\right]\, ,
\eea
where a prime denotes a derivative with respect to conformal time.
The second term is the source of the two-mode squeezing term while the first one is the usual free part of the quadratic Hamiltonian, as in Minkowski space.

In the Heisenberg picture, the equation for $a_{\bm k}$ is written
\begin{equation}\label{EoM}
 \frac{d a_{\bm k}}{d\tau} \, = \, \frac{z'}{z}a_{\bm k}^{\dagger} - ic_sk a_{\bm k}\,.
\end{equation}
In the case of inflation $\frac{z'}{z} = -\frac{1}{\tau}$. Hence, on super-Hubble scales (for which the momentum term on the right hand side of (\ref{EoM}) is negligible), the z term results in a highly squeezed vacuum. In contrast, during a phase of slow Ekpyrotic contraction $\frac{z'}{z} = -\frac{p}{(1-p)\tau}$ for super-Hubble modes. Since $p \ll 1$ the amount of squeezing is very small. In the limit $p \rightarrow 0$, the equation for the annihilation operators becomes the same as in flat spacetime quantum field theory.

Since the squeezing term is sub-dominant for Ekpyrosis, we can safely ignore the squeezing entropy which plays a crucial role for inflation, and focus on the entanglement entropy generated by the leading gravitational nonlinearities, i.e. by the cubic interaction term in the Hamitonian. This will be the topic of the following sections.

\section{Entanglement entropy for scalar perturbations due to gravitational non-linearities}

In this section we turn to the computation of the momentum space entanglement entropy between super- and sub-Hubble modes generated by the gravitational nonlinearities. In applications to black hole physics and in the context of the AdS/CFT correspondence, entanglement entropy is usually considered in terms of position space entanglement \cite{EE}. However, when considering the entropy of cosmological perturbations, it is more natural to consider momentum space entanglement. The reason is that the basic variables are the momentum modes of the fluctuations. The modes that classicalize and become accessible to cosmological experiments are the super-Hubble modes, with the sub-Hubble modes acting as a sea which are not directly accessible. Hence, it is natural to consider the entanglement between super- and sub-Hubble modes.  In a previous paper \cite{us}, momentum space entanglement entropy was studied in the context of inflation, generalizing the formalism developed in \cite{Bala} (see also \cite{previous, Shanki} for earlier work on the entropy of cosmological perturbations).

The entanglement entropy due to the mode coupling by the leading order cubic nonlinear terms can be calculated as follows. First, we break up the Hilbert space of scalar perturbations into sub-Hubble and super-Hubble modes, \textit{i.e.}
\be
\mathcal{H} \, = \, \mathcal{H}_{\rm sub} \otimes \mathcal{H}_{\rm super} \, .
\ee
The super-Hubble modes shall be treated as our system which is coupled to the bath of sub-Hubble modes. We shall carry out this calculation in Fourier space. Since our analysis is in the framework of an effective field theory, we need to apply $\Mpl$ as a physical cutoff for our sub-Hubble modes. The total Hamiltonian can be described as
\begin{eqnarray}
 {\bf H} \, = \, {\bf H}_{\rm sub} \otimes \mathbb{I} + \mathbb{I} \otimes {\bf H}_{\rm super} + {\bf H}_{\rm int}\,,
\end{eqnarray}
where ${\bf H}_{\rm sub}$ and ${\bf H}_{\rm super}$ refer to the quadratic Hamiltonian for the scalar modes with momenta $k < a H$ and $k > a H$, respectively. ${\bf H}_{\rm int}$ refers to the interaction Hamiltonian due to our cubic non-Gaussian term.

As discussed in \cite{us} (based on \cite{Adshead} and \cite{Malda}), in terms of the variable $\zeta$, the integrated interaction Lagrangian is given by
\begin{align}\label{CubicLag}
 S_3 \, = \, \Mpl^2 \int \d t\, \d^3 x & \left[a^3\epsilon^2 \zeta\dot{\zeta}^2 + a\epsilon^2\zeta(\partial\zeta)^2 \right. \nonumber    \\
                                       & \left.-\frac{\d}{\d t}\left(a^3\epsilon^2\mathcal{\zeta}^2\dot{\mathcal{\zeta}}\right)\right]\, 
\end{align}
where we neglected nonlocal terms and used the fact that $\epsilon$ is constant. Since in the case of the Ekpyrotic scenario the dominant mode of $\zeta$ is constant, the leading interaction term in the action is
\be\label{CubicLag2}
S_3 \, = \, \Mpl^2 \int \d t\, \d^3 x  a\epsilon^2\zeta(\partial\zeta)^2 \, .
\ee
Using conformal time, and in terms of the canonical variable $v$, this interaction term takes the form
\be\label{S3}
S_3 \, = \,   \frac{\sqrt{\epsilon}}{2\sqrt{2}\,a\, \Mpl}\int \d\tau\, \d^3{\bf x}\, v\,(\partial v)^2
\ee

To make contact with the formalism to compute the entanglement entropy developed in \cite{Bala} and applied in \cite{us} to the case of inflationary cosmology, we note that the effective coupling $\lambda$ is
\be\label{lambda}
\lambda \, = \, \frac{\sqrt{\epsilon}}{2\sqrt{2}a \Mpl} \, ,
\ee
from which we can define a dimensionless interaction parameter, given by
\be\label{lambda1}
\tilde{\lambda} \, = \, \frac{\sqrt{\epsilon} \Lambda}{2\sqrt{2}a \Mpl} \, ,
\ee
where $\Lambda$ is a renormalization scale which we expect to be the Planck scale.

\section{Entanglement Entropy of scalar perturbations}

As a consequence of the nonlinearities, an initial vacuum state of both system and bath modes gets excited. At time $t$ the state $|\Omega\rangle$ becomes
\bea
|\Omega\rangle(t) \, = \, |0,0\rangle &+& \sum_{n\neq 0} A_n(t) |n,0\rangle + \sum_{n\neq 0} B_N(t)  |0,N\rangle \nonumber \\
&+&\, \sum_{n,N\neq 0} C_{n,N}(t) |n,N\rangle\, ,
\eea
where $|n\rangle$ denotes an n-particle state of the system (in fact, a product state over all super-Hubble $k$ modes), and $|N\rangle$ is the corresponding state for the bath. At the initial time, the coefficients $A_n, B_N$ and $C_{n, N}$ all vanish, and they build up gradually over time due to the gravitational interactions.

As shown in \cite{Bala} and generalized in \cite{us} in the case of a time-dependent background, the induced entanglement entropy between the super- and sub-Hubble modes is induced by the interaction coefficients $C_{n, N}$ in the following way
\be\label{Ent}
S_{\rm ent} \, = \,  - \lambda^2 \log\(\tilde{\lambda}^2\) \sum_{n,N\neq 0} |\tilde{C}_{n,N}|^2\,.
\ee
For an infinite set of modes labelled by a continuous index $k$, the summation becomes a momentum space integral. We use the dimensionless effective coupling as the argument of the logarithm to make this term well-defined. However, in the end, this term will not be too important for our purposes as we shall ignore the logarithm term to focus only on the rest of the factors to get an order of magnitude estimate for the bound on the energy scale of the bounce.

We follow the analyses of \cite{us} in order to calculate the entanglement entropy in this case. For the contracting phase in Ekpyrosis, the squeezing of the state of fluctuations on super-Hubble scale is negligible and we can set $r_k(\eta) \sim 0$ for the calculation. Hence, the general formula (62) of \cite{us} simplifies to yield the following expression for the induced entanglement entropy density per comoving volume (please see the Appendix for details)
\begin{widetext}
 \begin{align}
  s \, = \, & (2\pi)^3\lambda^2 \log\(\tilde{\lambda}^2\) \int_{k_I}^{aH} \d^3 p_3\, \int_{aH}^{a\Mpl}\d^3p_2\, \int_{aH}^{a\Mpl}\d^3 p_1\, \delta^3(p_1+p_2+p_3)\,\frac{1}{(p_1+p_2+p_3)^2} \(\frac{p_1p_2}{p_3}\)\label{eq: Entag_Inte}                                                          \\
  =         & -\lambda^2 \log\(\tilde{\lambda}^2\)\int^{aH}_{k_I}\frac{\d^3p_3}{(2\pi)^3}                       \int_{aH}^{aM_{pl}}\frac{\d^3 p_2}{(2\pi)^3} \bigg(\frac{p_2\sqrt{p_2^2+p_3^2+2p_2p_3\cos\theta}}{p_3}\bigg)\frac{1}{(\sqrt{p_2^2+p_3^2+2p_2p_3\cos\theta} + p_2+p_3)^2} \nonumber 
 \end{align}
\end{widetext}
Notice that the super-Hubble mode $p_3$ gets an infrared cut-off by $k_I$ while the sub-Hubble mode $p_2$ has a natural, physical UV-cutoff given by $aM_{pl}$. The meaning of the IR cut-off can be thought of as follows: If there is a phase prior to Ekpyrosis, then we shall only consider super-Hubble modes generated during Ekpyrosis for our calculations. However, note that unlike in the case of inflation, we are free to take the limit $k_I \rightarrow 0$ for our integral without encountering any divergences. Generally we take $p_3\ll p_2$ when \eqref{eq: Entag_Inte} is evaluated. Thus, we obtain
\begin{align} \label{result}
 s \, \approx \, & \frac{\lambda^2 \log\left(\tilde{\lambda}^2\right)}{4\pi^4}\int^{aH}_{k_I}\d p_3\,                                                 
 p_3^2 \int^{a\Mpl}_{aH} \d p_2\, p_2^2 \frac{1}{4p_3}\nonumber         \\
 \approx   \,    & \frac{\lambda^2 \log\left(\tilde{\lambda}^2\right)}{48\pi^4}\bigg[(aH)^2-k_I^2 \bigg]\cdot \bigg[(a\Mpl)^3-(aH)^3 \bigg] \nonumber \\
 \sim   \,       & \frac{1}{192\pi^4 p}  \log\left(\frac{1}{2\sqrt{2}p a}\right) a^3 \Mpl H^2                                            
\, . 
\end{align}
In the last line, we have replaced the slow-roll parameter $\epsilon$ by the small number $p$ and set $\Lambda = \Mpl$. Dividing by the factor $a^3$ we obtain the the entropy density per physical volume element. We see that the entanglement entropy (per unit physical volume) grows logarithmically as a function of time in the contracting phase, ignoring the time-dependence of $H$.

If we interpret entanglement entropy as a contribution to the entropy which should obey the second law of thermodynamics, then we can obtain an upper bound on the energy scale of the bounce (or equivalently on the value of $H$ just before the bounce) by demanding that
the entanglement entropy (\ref{result}) be smaller than the thermal entropy density after the bounce which is calculated as follows

\begin{equation}
 \frac{s}{V_0} = \frac{4}{3}\big(\frac{30}{\pi^2 g}\big)^{1/4}\rho^{1/4}
 =  \frac{4}{3}\big(\frac{90}{\pi^2 g}\big)^{1/4}H^{3/2}M_{pl}^{3/2}
 \label{eq:thermal entropy}
\end{equation}
where $g$ is the number of effective spin degrees of freedom in the thermal bath after the bounce. From \eqref{eq:thermal entropy} and \eqref{result} we obtain the condition
\be \label{bound}
\left( \frac{H}{\Mpl} \right)^{1/2} \, < \, 256 \pi^{7/2} \big(\frac{90}{g}\big)^{1/4} p \, .
\ee
This sets an upper bound on the string scale, the energy scale where the S-brane will occur. It is not a very stringent bound. Assuming that the S-brane arises at string energy density, then the above equation yields the corresponding bound on the string energy scale in Planck units.
\\

\section{Conclusions}

We have computed the entanglement entropy between sub- and super-Hubble modes during a phase of Ekpyrotic contraction. We found that this entropy grows logarithmically with decreasing scale factor. Demanding that the entanglement entropy at the end of the phase of contraction does not exceed the thermal entropy after the bounce, we obtain an upper bound on the energy scale of the bounce.

\section*{Acknowledgments}

The research at McGill is supported in part by funds from NSERC and from the Canada Research Chair program. SB is supported in part by a McGill Space Institute fellowship and by a CITA National fellowship. ZW is supported in part by funds from a Templeton foundation sub-contract.

\section*{Appendix: Calculation of entanglement entropy}

In order to explicitly calculate the entanglement entropy, we first need to evaluate the matrix elements, given by
\begin{eqnarray}\label{matrix}
 C_{n,N} := -i\int_{\tau_0}^{\tau} \d\tau'\, e^{i\(p_1+p_2+p_3\) \tau'} \, \left\langle n,N\right| H_{\rm int}(\tau') \left|0,0\right\rangle\,.
\end{eqnarray}
Note that the interaction Hamiltonian can be derived from \eqref{S3} and is given by
\begin{widetext}
 \begin{eqnarray}\label{Hint}
  \lambda(\tau)	H_{\rm int} &=& \lambda(\tau) \int_{\Delta}  \left[ \sqrt{\frac{k_2 k_3}{k_1}} \left(c^\dagger_{-\textbf{k}_1} c^\dagger_{-\textbf{k}_2} c^\dagger_{-\textbf{k}_3} + c_{\textbf{k}_1} c^\dagger_{-\textbf{k}_2} c^\dagger_{-\textbf{k}_3} +  \ldots \right) + \sqrt{\frac{k_2 k_1}{k_3}} \left(c^\dagger_{-\textbf{k}_1} c^\dagger_{-\textbf{k}_2} c^\dagger_{-\textbf{k}_3} +\ldots \right)\right. \nonumber\\
  && \hspace{2cm} \left.+ \sqrt{\frac{k_1 k_3}{k_2}} \left(c^\dagger_{-\textbf{k}_1} c^\dagger_{-\textbf{k}_2} c^\dagger_{-\textbf{k}_3} + \ldots\right) \right]\,,
 \end{eqnarray}
\end{widetext}
where $\lambda$ is given by \eqref{lambda} and we have defined $\int_\Delta := \int \frac{\d^3 k_1}{(2\pi)^3}\,\frac{\d^3 k_2}{(2\pi)^3}\,\frac{\d^3 k_3}{(2\pi)^3}\, \(2\pi\)^3 \delta^3(\textbf{k}_1+\textbf{k}_2+\textbf{k}_3)$ and the `dots' are permutations.

Since the squeezing is negligible for Ekpyrosis, both the sub- and super-Hubble modes are in the Minkowski vacuum. This makes the evaluation of $\left\langle n,N\right| H_{\rm int}(\tau') \left|0,0\right\rangle$ relatively simpler compared to the case for inflation, and we need to only consider terms in $H_{\rm int}$ of the form $c^\dagger_{-\textbf{k}} c^\dagger_{-\textbf{k}} c^\dagger_{-\textbf{k}}$, where two of the modes can be sub-Hubble and one super-Hubble or the other way around. Once again, since the vacuum state is the same for both sub- and super-Hubble modes, the dominant contribution is going to be from the case where there are two super Hubble modes. We find:
\be
 \left\langle n,N\right| H_{\rm int}(\tau') \left|0,0\right\rangle \sim  \sqrt{\dfrac{p_1 p_2}{p_3}}\,,
\ee
where $p_1, p_2$ are sub-Hubble and $p_3$ is the super-Hubble mode. (The other terms proportional to $\sqrt{p_1 p_3/p_2}$ and $\sqrt{p_2 p_3/p_1}$ are sub-dominant in this case.) Plugging this into the expression for the matrix element \eqref{matrix}, we find
\begin{eqnarray}
 C_{n,N} = -i\int_{\tau_0}^{\eta} \d\tau'\, e^{i\(p_1+p_2+p_3\) \tau'} \,  \sqrt{\dfrac{p_1 p_2}{p_3}} \,.
\end{eqnarray}
This is where the negligible squeezing for Ekpyrosis plays a crucial role so that we can evaluate the matrix elements as if for \textit{time-independent} perturbation theory. Using this, one can calculate the entanglement entropy as
\begin{widetext}
 \begin{eqnarray}
  S_{\rm ent} &\sim& (2\pi)^3\;\; \int_H^{a H}\dfrac{\d^3p_3}{(2\pi)^3} \int_{aH}^{a\Mpl}\dfrac{\d^3p_2}{(2\pi)^3} \int_{a H}^{a\Mpl} \dfrac{\d^3p_1}{(2\pi)^3}\;\delta^3(\textbf{p}_1 + \textbf{p}_2 + \textbf{p}_3) \(\frac{p_1 p_2}{p_3}\)\times\nn\\
  && \hspace{4cm}\left|\int_{\tau_0}^\tau\,\d\tau'\; e^{i\(p_1 + p_2 + p_3\) \eta'}\right|^2\,\lambda^2(\tau') \log\(\tilde{\lambda}^2(\tau')\) \; \,.
 \end{eqnarray}
\end{widetext}
The time-dependence of the coupling parameter $\lambda(\tau)$ is extremely weak in Ekpyrosis since it is only due to the presence of the scale factor $a(\tau)$, which itself changes very slowly in conformal time. Therefore, one can pull the $\lambda^2 \log\(\tilde{\lambda}^2\)$ term outside the integral and then one arrives at the result given in \eqref{Ent}, with $\tilde{C}_{n,N}$ being the time-independent matrix elements. As shown in this Appendix, in the case of Ekpyrosis, one can easily work with the time-independent matrix elements (due to the weak dependence of the interaction parameter on conformal time and having an usual flat vacuum for the super-Hubble modes) and we recover the results discussed in the main body of the text.

\end{document}